\documentclass[floatfix,aps,prl,reprint]{revtex4-1}

\usepackage{graphicx}
\usepackage{dcolumn}
\usepackage{bm}

\usepackage[utf8]{inputenc}
\usepackage[T1]{fontenc}
\usepackage{mathptmx}

\usepackage{epstopdf}

\draft 

\begin{document}


\title{Femtosecond laser-induced electron emission from nanodiamond-coated tungsten needle tips} 



\author{A. Tafel}
\email[]{alexander.tafel@fau.de}
\affiliation{Department of Physics, Friedrich-Alexander-Universit\"at Erlangen-N\"urnberg, Staudtstra\ss{}e 1, D-91058 Erlangen, Germany}

\author{S. Meier}
\affiliation{Department of Physics, Friedrich-Alexander-Universit\"at Erlangen-N\"urnberg, Staudtstra\ss{}e 1, D-91058 Erlangen, Germany}

\author{J. Ristein}
\affiliation{Department of Physics, Friedrich-Alexander-Universit\"at Erlangen-N\"urnberg, Staudtstra\ss{}e 1, D-91058 Erlangen, Germany}

\author{P. Hommelhoff}
\affiliation{Department of Physics, Friedrich-Alexander-Universit\"at Erlangen-N\"urnberg, Staudtstra\ss{}e 1, D-91058 Erlangen, Germany}


\date{\today}

\begin{abstract}
We present femtosecond laser-induced electron emission from nanodiamond-coated tungsten tips. Based on the shortness of the femtosecond laser pulses, electrons can be photo-excited for wavelengths from the infrared (1932 nm) to the ultraviolet (235 nm) because multi-photon excitation becomes efficient over the entire spectral range. Depending on the laser wavelength, we find different dominant emission channels identified by the number of photons needed to emit electrons. Based on the band alignment between tungsten and nanodiamond, the relevant emission channels can be identified as specific transitions in diamond and its graphitic boundaries. It is the combination of the character of initial and final states (i.e. bulk or surface-near, direct or indirect excitation in the diamond band structure), the number of photons providing the excitation energy and the peak intensity of the laser pulses that determines the dominant excitation channel for photo-emission. A specific feature of the hydrogen-terminated nanodiamond coating is its negative electron affinity that significantly lowers the work function and enables efficient emission from the conduction band minimum into vacuum without energy barrier. Emission is stable for bunch charges up to 400 electrons per laser pulse. We infer a normalized emittance of < 0.20~nm~rad and a normalized peak brightness of > $1.2 \cdot 10^{12}~\text{A}~\text{ m}^{-2}~\text{ sr}^{-1}$. The properties of these tips are encouraging for their use as laser-triggered electron sources in applications such as ultrafast electron microscopy and diffraction and novel photonics-based laser accelerators.
\end{abstract}


\maketitle 

Tip-shaped cathodes are amongst the most commonly used electron sources in electron microscopy due to their ability to provide a high quality beam. Typical materials are zirconia in common Schottky type emitters and lanthanum hexaboride because of their low work function, and tungsten due to the easy fabrication of sharp tips ideally suited for (cold) field emission \cite{Spence2013}. Most of the commonly used emitters are operated under ultra-high vacuum conditions in the $10^{-8} - 10^{-9}$~Pa regime to minimize bombardment with ionized gas and adsorption on the emitter surface. Furthermore, they are heated for thermal enhancement of the emission or to achieve stable operation due to reduced adsorption. \\
Over the last decades, ultrafast electron microscopy has emerged \cite{Bostanjoglo2000,Zewail2010_Review,Arbouet2018_UTEM_review}. Until today, the same emitters are used in DC and ultrafast mode. In the latter case, the cathode is typically triggered by femtosecond laser pulses resulting in femto- to picosecond electron pulses \cite{Yang2010,Plemmons2015_LaB6_UTEM,Sun2015,Feist2017,Houdellier2018,Kozak2018}. One of the major drawbacks of these laser-triggered electron sources is the continuous decrease of emission current over time \cite{Yang2010,Houdellier2018,Kozak2018}, which is attributed to laser-induced changes at the emitter surface. Femtosecond laser-induced photo-emission from tip-shaped cathodes has been extensively studied for the materials tungsten \cite{Hommelhoff2006,Hommelhoff2006PRL_tungsten,Barwick2007_tungsten,Krueger2011_Nature_tungsten,Bionta2013_tungsten_ATI,Meier2018}, gold \cite{Ropers2007_PRL_gold,Bormann2010_PRL_gold,Wimmer2014_THz_gold}, silver \cite{Bionta2016_silver}, hafnium carbide \cite{Kealhofer2012_HfC} and carbon nanotubes \cite{Bionta2015_CNT,Li2017_CNT}. Pulsed photo-emission from single crystal diamond tips has been investigated with nanosecond pulses \cite{Porshyn2017_diamond_ns_laser}. Femtosecond photo-emission from tip-shaped heterostructures offers promising opportunities yet to be discovered.\\
Diamond is one of the most robust materials due to its exceptional chemical inertness, mechanical strength and thermal conductivity. Nanocrystalline diamond (NCD) is a good electron emitter, especially if the surface exhibits negative electron affinity (NEA) \cite{Cui2000_JAP}. The graphitic grain boundaries in this composite material provide electrical conductivity, and the low work function of the diamond matrix that goes along with the NEA also lowers the surface energy barrier for the electrons even if they originate from the graphitic parts \cite{Cui2000_JAP}. NEA is also known to boost the photo-electron yield due to fundamental absorption, i.e. optical excitation across the band gap: electrons photo-excited into the diamond conduction band can be emitted into vacuum without any barrier after migration to the surface \cite{Himpsel1979_NEA_yield,Cui1999_NEA_photoemission}. The electron affinity of hydrogen-terminated diamond is as low as -1.3~eV for both main crystallographic surfaces (100) \cite{Maier_2001} and (111) \cite{Cui1998PRL}. \\
The combination of high beam quality from tip-shaped photocathodes with the mechanical strength and low work function of hydrogen-terminated diamond promises a robust and high-brightness photocathode. Here we present first photo-emission results from a tip-shaped semiconductor/metal heterostructure - diamond-coated tungsten tips - triggered with femtosecond laser pulses and characterize the underlying photo-emission physics by identifying various emission channels. We define an emission channel as the combination of photon energy and energy states involved in the photo-emission process of electrons.

\begin{figure}
\centering
\includegraphics[width=8 cm   ]{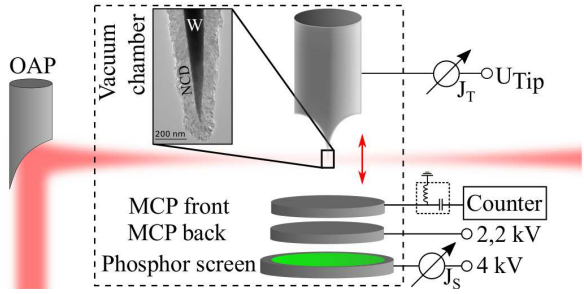}
\caption{\label{fig:setup} 
Experimental setup. The laser is focused onto the diamond-coated tungsten tip with an off-axis parabolic mirror (OAP). The inset shows a transmission electron micrograph of the nanodiamond-coated tungsten tip. A voltage clearly below the DC field emission threshold is applied between tip and microchannel plate (MCP) to accelerate electrons towards the MCP. See text for details.}
\end{figure} 

To obtain the nanodiamond-coated tips, 100 $\mu$m diameter tungsten wire is electrochemically etched resulting in a tip with a typical apex radius of roughly 10 nm. The freshly etched tip is dip-seeded in nanodiamond suspension and dry-blown with pressurized nitrogen. NCD is grown on the seeded tips with microwave-enhanced chemical vapor deposition resulting in a dense film of hydrogen-terminated nanocrystalline diamond with negative electron affinity covering the tungsten surface (Fig. \ref{fig:setup} inset). A thin layer of tungsten carbide (WC) is expected to be formed at the diamond/tungsten interface \cite{Davidson1978}. Samples used in this work have apex radii between 60~nm and 200~nm including the diamond coating. Details of the fabrication process and a structural characterization of the tips are published elsewhere \cite{Tafel2019_DRM}.

The so-fabricated tips are mounted in an ultra-high vacuum chamber with a base pressure of $1 \cdot 10^{-7}$~Pa. Femtosecond laser pulses are focused at the tip with the help of a 51 mm diameter off-axis parabolic mirror with 152 mm focal length outside of the vacuum chamber, resulting in a measured spot radius of 3.8 $\mu$m  at 512 nm ($1/e^2$ intensity radius). The employed commercial laser system consists of a regeneratively amplified Ti:Sa oscillator (1 kHz repetition rate, 80 fs pulse duration), an optical parametric amplifier and a stage for second harmonic and sum-frequency generation. Additionally, a Ti:Sa oscillator (780 nm, 80 MHz, 6 fs) is used for long-term stability measurements. We apply a negative voltage below 50 \% of the DC field emission threshold (400-2000~V depending on the individual tip). Due to the dielectric surface with a small work function of 2.8~eV (equation \ref{eq:work_function}), the Schottky reduction is lower compared to metal tips and is neglected here. The DC field was chosen low enough, so that photon-assisted field emission does not occur, only multi-photon emission. The laser pulses are linearly polarized parallel to the tip axis. Photo-emitted electrons are detected with a multichannel plate (MCP) with grounded front plate. For bunch charges below one electron per laser pulse, we count detection events on the MCP, above one electron per pulse we measure the calibrated MCP screen current and for large average currents at high repetition rates we are able to additionally measure the current through the tip.

\begin{figure}
\centering
\includegraphics[width=8 cm   ]{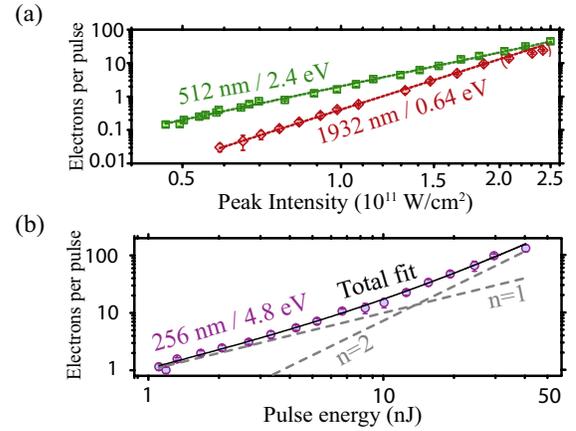}

\caption{\label{fig:256_512_1932} 
(a) Photo-emission at 1932~nm (red diamonds) and 512~nm (green squares), with slopes 5.0 (red dashed line, last three data points not included due to potential saturation effects) and 3.4 (green dashed line). (b) Transition from one- to two-photon emission at 256~nm. The grey dashed lines are the corresponding contributions, the solid black line is the sum of the two contributions. Note that we used pulse energy instead of peak intensity in (b) as we could not measure the laser spot size in the UV.}
\end{figure}

\begin{figure}
\centering
\includegraphics[width=8 cm   ]{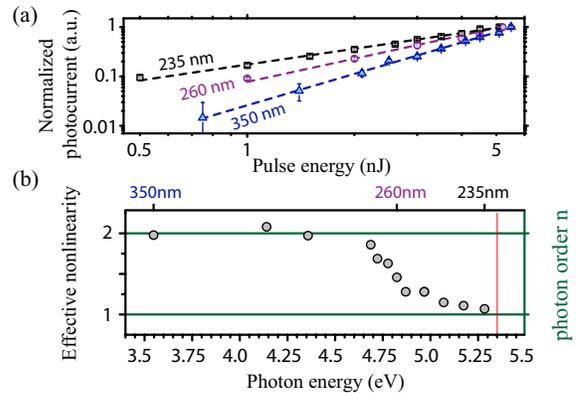}
\caption{\label{fig:UV-emission} 
Power scaling in the UV with transition from one- to two-photon emission as the dominant channel. (a) Data at 235 (black squares), 260 (violet circles) and 350~nm (blue triangles) with effective nonlinearities 1.1, 1.4 and 2.0, respectively. (b) Effective nonlinearity vs photon energy with a gradual transition from n=1 to n=2 at roughly 4.8~eV.}
\end{figure}

In order to identify the different contributions to the photo-electron current $J$, we have measured its dependence on the peak intensity $I_p$. Due to the high $I_p$ of the femtosecond laser pulses, optical excitation is not limited to one-photon absorption processes as multi-photon absorption becomes efficient. The dependence of the photo-electron current $J$ on $I_p$ is expected to be a sum of power law contributions:

\begin{equation}
    J = \sum_{n}^{\infty} a_n \cdot I_p^n
    \end{equation}

where $n$ reflects the number of photons necessary to provide the excitation energy and $a_n$ the corresponding coefficient for the specific emission channel. Often, one channel is dominant, hence the slope of log($J$) vs log($I_p$) directly reveals the photon order $n$. If more than one channel is involved with comparable strength, the linearized slope is non-integer and is called effective nonlinearity. Depending on the photon energy and laser intensity, different emission channels can become dominant. We show the power dependence of the photo-electron current at a wavelength of 1932, 512 and 256~nm in Fig. \ref{fig:256_512_1932}. For 1932~nm, we find an integer slope of 5.0 indicating a single dominating emission channel with 5 photons. At 512~nm, the plot shows an effective nonlinearity of 3.4. This is indicative for two channels with photon order 3 and 4 contributing. At 256~nm, we observe a transition from photon order one at low intensities to photon order two at high intensities. \\
In the UV (235-350~nm) we have investigated the wavelength dependence of the effective nonlinearity in more detail. Fig. \ref{fig:UV-emission} (a) shows the log($J$) vs log($I_p$) plot for 235, 260 and 350~nm. We find effective nonlinearities of 1.1, 1.4 and 2.0, respectively. Again, this reflects the transition of the dominant emission channel from first to second order. Note that we do not observe a transition in the power dependence directly in contrast to Fig. \ref{fig:256_512_1932} (b). This is due to the restricted pulse energy range in Fig. \ref{fig:UV-emission}. The effective nonlinearities for all wavelengths in the UV are summarized in Fig. \ref{fig:UV-emission} (b), confirming the transition mentioned above.

\begin{figure}
\centering
\includegraphics[width=8 cm   ]{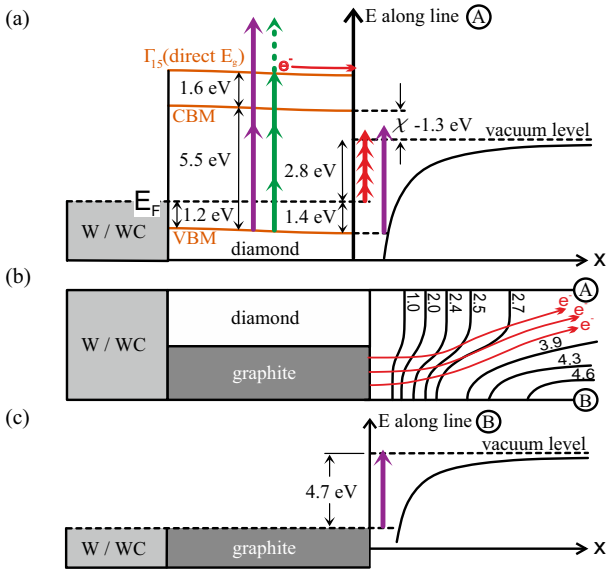}
\caption{\label{fig:emission_scheme} Sketch of the relevant energy levels and optical excitation paths in the nanodiamond needle tip coating, including the graphitic boundaries. The concatenated arrows indicate excitation channels we identify as relevant here. Length and color of the arrows in (a) and (c) represent the photon energies 0.64 (near infrared), 2.4 (green) and 4.8~eV (near ultraviolet). With these photo-excitation channels, we can explain the observed laser power and wavelength dependence discussed around Figs. 2 and 3. Intriguingly, this emission channel identification, except for the assignment of the one photon process in the UV, seems to result in a unique attribution in spite of the intricate level structure. The work function of diamond is 2.8~eV for a negative electron affinity of $\chi$ = -1.3~eV. In the diamond bulk, electrons can be excited across the indirect (5.5 eV)  or direct bandgap (7.1 eV), with two ultraviolet or three green photons, migrate to the surface and cross the surface into vacuum. Alternatively, electrons can be excited into vacuum directly by one ultraviolet or five infrared photons. Even if the electrons originate from graphite close to the diamond interface, they effectively feel the work function of diamond as indicated by their trajectories across the equipotential lines in (b). See text for details.}
\end{figure}

For the interpretation of the data, we sketch the energy states relevant for this work in Fig. \ref{fig:emission_scheme}. Five junctions between W/WC, diamond, vacuum and the graphitic grain boundaries (called graphite in Fig. \ref{fig:emission_scheme}) are formed. Diamond forms Schottky junctions with graphite and W/WC with Schottky barriers $E_{B,G}$=1.4~eV \cite{Cui1999} and $E_{B,W/WC}$=1.2~eV \cite{shiomi1989_W_diamond}, respectively. As the sample surface only consists of diamond grains and their graphitic boundaries, the junctions diamond/vacuum and graphite/vacuum are the relevant ones for electron emission into vacuum. In a hetero system involving metallic (W/WC and in a good approximation also the half metal graphite) and semiconducting (diamond) components, the Fermi level in the semiconductor relative to the valence band maximum (VBM) is identical to the Schottky barrier height as long as the dimensions of the semiconducting parts are much below the Debye length of the semiconductor. This is certainly the case for the diamond grains. We expect $E_{B,G}$ to dominate at the diamond surface as the average grain size (approx. 20~nm) is smaller than the thickness of the diamond film. Consequently, the Fermi level is $E_{B,W/WC}$=1.2~eV above the VBM at the back contact and $E_{B,G}$ =1.4~eV above the VBM at the free surface (see Fig. \ref{fig:emission_scheme}(a)). The work function $\Phi$ is defined as the energy difference between the vacuum level and the surface Fermi level. Graphite has a work function of 4.7~eV \cite{Willis1974}, while the work function of diamond depends on the electron affinity $\chi$ and $E_{B,G}$ and results in

\begin{equation} \label{eq:work_function}
    \Phi_{dia} = E_g - E_{B,G} + \chi  = (5.5-1.4-1.3) \text{ eV} = 2.8 \text{ eV},
\end{equation}

where we inserted -1.3~eV for the electron affinity of a fully hydrogen-terminated diamond surface \cite{Cui1998PRL,Cui_2000}. The diamond work function also constitutes the low energy threshold for electrons originating from graphite (see Fig. \ref{fig:emission_scheme}(b)). \\
Based on this band diagram, we can identify electron emission channels with different energy thresholds as indicated in Fig. \ref{fig:emission_scheme}(a). For diamond with negative electron affinity, the energy barrier between the conduction band minimum (CBM) at the surface does not exist: Electrons can be excited into the conduction band across the indirect (5.5~eV \cite{Dean1965_indirect_bandgap}) and direct bandgap (7.1~eV \cite{Giustino2010_PRL_direct_bandgap,Logothetidis1992_direct_bandgap}), migrate to the surface and escape straight into vacuum even if they have thermalized to CBM. Alternatively, direct optical excitation from electronic states at the surface to the plane wave like states in vacuum can lead to photoelectrons as well. 

The emission probabilities of the different channels are complex functions of the densities of initial and final states, the number of photons necessary to provide the transition energy and the laser intensity. We discuss them by referring to their signature in the log($J$) vs log($I_p$) plots in Figs. \ref{fig:256_512_1932} and \ref{fig:UV-emission}. 
Excitation with 1932~nm ($\hbar \omega$~=~0.64~eV, red arrows in Fig. \ref{fig:emission_scheme}(a)) and an observed photon order 5 can be identified as transitions at the surface from the Fermi level to the vacuum level. For clearer presentation, we have sketched the red arrows only in Fig. \ref{fig:emission_scheme}(a) although the initial states at the Fermi level can be assigned either to defects in the diamond or, more likely, to the graphitic grain boundaries \cite{Cui1999_NEA_photoemission}. \\
At 512~nm ($\hbar \omega$~=~2.4~eV, green arrows in Fig. \ref{fig:emission_scheme}) the effective nonlinearity equals 3.4, which we attribute to excitation across the direct band gap by three or by four photons. \\
With UV excitation ($\lambda$~<~350~nm, $\hbar \omega$~>~3.5~eV, violet arrows in Fig. \ref{fig:emission_scheme}), we observe one- and two-photon processes (Fig. \ref{fig:256_512_1932}(b) and \ref{fig:UV-emission}). We assign the one photon process at low intensities to excitation at the surface from the Fermi level or the diamond VBM to the vacuum level. Evaluating energy differences only, the two-photon process ($\Delta E$~>~7.2~eV for $\lambda$~<~350~nm) could be assigned to all transitions in the band diagram of Fig. \ref{fig:emission_scheme}. We suggest for this process the transition across the direct band gap of diamond: The spatial overlap of the wave functions, the direct nature of the transition and the large excitation volume make this process by far the most likely. This argument is supported also by the nonlinearity of 3.4 which we observe for 512~nm ($\hbar \omega$~=~2.4~eV: Two photons of that energy would suffice to excite electrons from the VBM directly into vacuum. Nevertheless, this channel is not observed as the dominant one. The situation is different for 1932~nm ($\hbar \omega$~=~0.64~eV): With this wavelength excitation across the indirect bandgap would require 9 photons and 11 photons across the direct bandgap. These extremely high order processes are so unlikely that we observe the fifth order process at the surface as the dominant channel instead. For the spectral range investigated, we find no evidence of absorption across the indirect bandgap as the dominant mechanism. \\

Last, we characterize the photo-emission stability over time at different bunch charges and estimate the normalized peak brightness $B_{p,norm}$. For best comparison of $B_{p,norm}$ with existing literature on ultrafast tip-shaped electron sources \cite{Feist2017,Houdellier2018}, we calculate all quantities as normalized root-mean-squared (rms) values and use the following definition: 
\begin{equation}
    B_{p,norm} = \frac{J_p}{4 \cdot \pi^2 \cdot \epsilon_{x,norm} \cdot \epsilon_{y,norm}},
\end{equation}

where $J_p$ is the peak current, $\epsilon_i$ are the transverse emittances and the subscript $norm$ indicates normalized values. As an upper bound for the transverse emittances, we measure the emission angles $\alpha_i$ and assume homogeneous emission across the geometrical radius of the emitter ($r$=170~nm, $r_{rms}=\frac{r}{\sqrt{3}}$). Note that the effective source size and therefore the emittance of tip-shaped emitters can be an order of magnitude smaller as the curved surface induces correlations between origin and transverse momentum \cite{Ehberger2015,Meier2018}. Photo-emission at 1932~nm and 40~eV electron energy yields $\alpha_x = 0.16(6)$~rad, $\alpha_y = 0.15(9)$~rad, $\epsilon_{x,norm} = 0.20$~nm~rad and $\epsilon_{y,norm} = 0.19$~ nm~rad. Assuming that the emission duration matches the laser pulse duration, we calculate the normalized rms peak brightness $B_{p, norm} = 1.2~\cdot~10^{12}~\text{A}~\text{ m}^{-2}~\text{ sr}^{-1}$ for one electron per pulse, comparable to a femtosecond cold field emitter at 15~electrons per pulse \cite{Houdellier2018}. Because we use the geometrical and not the effective source size and because we consider currents of one electron per pulse, we consider this peak brightness a lower bound. 

\begin{figure}
\centering
\includegraphics[width=8 cm   ]{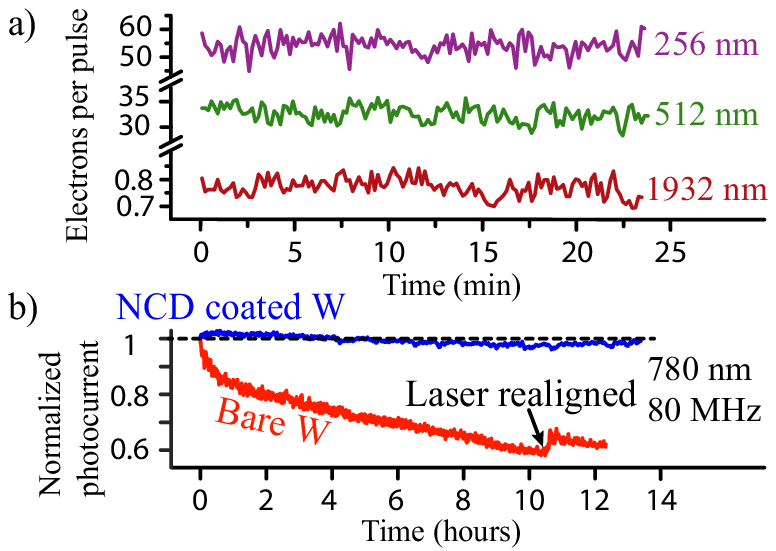}
\caption{\label{fig:stability} a) Photo-emission at 1 kHz repetition rate is stable at 256, 512 and 1932~nm from the UV to the infrared both below and above one electron per pulse. The noise is caused by laser power and pointing fluctuations. \\
b) Long-term photo-emission at 80 MHz and 780 nm is stable from a diamond-coated tungsten tip (blue) and unstable from a bare tungsten tip (red). The short term fluctuations (over 1 min) from the coated and uncoated tip are 3\% and 5\%, the bunch charges at t=0 are 25 and 6.5 electrons per pulse, respectively.}
\end{figure}
The photo-emitted current is stable over a time scale of at least half an hour at 256, 512 and 1932~nm with bunch charges of 55, 32 and 0.75~electrons per pulse, respectively (Fig. \ref{fig:stability}). With a stable 80~MHz Ti:Sa oscillator, the photo-current is stable over more than 12 hours and trillions of pulses. In contrast, the photo-emission from an uncoated monocrystalline [310]-oriented tungsten tip decays over time (Fig. \ref{fig:stability}); a comparable behavior with even stronger decay was observed in a transmission electron microscope ($\hbar \omega$~=~2.4~eV, p~=~$1~\cdot~10^{-9}$~Pa \cite{Houdellier2018}). Schottky emitters in scanning electron microscopes ($\hbar \omega$~=~3.6~eV, reduced barrier height $\Phi_{\mathrm{eff}}$~=~1.6~eV \cite{Yang2010} and $\hbar \omega$~=4.7~eV,  $\Phi_{\mathrm{eff}}$~=~2.8-3~eV, p~$ <4 \cdot 10^{-8}$~Pa \cite{Kozak2018}) show a similar behavior. Hence, nanodiamond-coated tungsten tips are more stable than these emitters, especially at low photon energies. (Working with low photon energies can be advantageous as the field enhancement at the apex \cite{Thomas2015} in combination with the nonlinearity enhances forward emission.) \\
In DC field emission occasional jumps occur, typical for cold field emission. The angular distribution in this emission mode is even smaller compared to laser-induced emission. \\  
We did not observe a change in laser-induced emission behaviour during our experiments with a laser fluence up to $30 \frac{mJ}{cm^2}$ and $3.4 \cdot 10^{11} \frac{W}{cm^2}$ peak intensity. Hence we find these as lower bounds of the damage threshold for diamond-coated tungsten tips. With 1932~nm pulses at $3.4~\cdot~10^{11}~\frac{W}{cm^2}$, we have measured 400~electrons per pulse. At these large bunch charges, pulse broadening due to Coulomb repulsion is expected to be severe \cite{Cook2016,Kozak2018}, which is why we have focused on smaller bunch charges.

In conclusion, we have presented femtosecond laser-induced electron emission from diamond-coated tungsten tips at 235-350~nm, 512~nm, 780~nm and 1932~nm. Based on the involved junctions between tungsten, diamond and the graphitic grain boundaries, we have proposed an emission model which explains our experimental data well. Individual emission channels can be selected by proper choice of laser intensity and wavelength. These channels are identified by the number of photons needed to emit an electron. Stable photo-electron current and the high brightness of the emitted electrons are encouraging to further investigate diamond-coated tungsten tips as an ultrafast electron source. \\
Before resubmission of this manuscript, we became aware of new and related work \cite{Borz2019}.

\section{Acknowledgement}

The authors acknowledge Mingjian Wu and Erdmann Spiecker for the transmission electron micrograph, funding from the Deutsche Forschungsgemeinschaft via grant SFB 953 "Synthetic Carbon Allotropes", from the European Research Council through grant “Near Field Atto” and the Gordon and Betty Moore Foundation via Grant GBMF4744 “Accelerator on a Chip International Program – ACHIP”.

\bibliography{bibliography_for_PRL.bib}

\end{document}